\documentstyle[twoside,epsfig]{article}
\date{}
\begin{document}
\section*{}
\setcounter{page}{0}
\newpage
\title{Neutron Scattering Studies of Heavy Fermion Systems}
\author{{\it By} T.E. Mason\\
\small{Department of Physics, University of Toronto,}\\
\small{Toronto, ON, Canada M5S 1A7}\\
{\it and} G. Aeppli\\
\small{NEC Research Institute,}\\
\small{4 Independence Way, Princeton, NJ, U.S.A. 08540}}

\maketitle
\begin{center}
\section*{\large {\bf Synopsis}}
\end{center}
\thispagestyle{myheadings}
{\footnotesize
\noindent
We review the results of recent studies of the elastic and
inelastic neutron scattering from a variety of heavy fermion compounds.
This class of materials exhibits a rich variety of ground states:
antiferromagnetically ordered, superconducting, semiconducting, and 
paramagnetic.  Neutron scattering from single crystals 
and powders has been a productive
tool for probing the magnetic order and fluctuations in all four cases.
This review deals with work on $\rm{UPt_3}$, $\rm{UPd_2Al_3}$, 
$\rm{UNi_2Al_3}$, $\rm{UNi_4B}$, CeNiSn, $\rm{Y_{1-x}U_xPd_3}$, and 
$\rm{UCu_{5-x}Pd_x}$.}

\vspace{0.5cm}
\tableofcontents

\section{Introduction}
\subsection{Overview}
Heavy fermion compounds, typically alloys containing U or Ce, are 
characterised by the small energy scale 
associated with the hybridization of nearly localised f-electrons with
conduction electrons.  This small energy scale means that properties such
as band structure, which are normally not considered temperature dependent,
can vary with temperature and are sensitive to small perturbations.  This
sensitivity gives rise to a rich variety of low temperature states in these
materials; for a review see Grewe and Steglich (1991).

At high temperatures heavy fermion systems behave as Kondo lattices and the
unpaired f electrons have a local magnetic moment that 
interacts with the conduction electrons in the same way as an isolated
Kondo impurity in a metal.  As the temperature is lowered, however, the 
magnetic moments no longer behave as isolated, localised impurities and the 
system enters the coherent state which is characterised by the large effective
mass (and enhanced electronic specific heat) associated with the 
quasiparticles of a strongly interacting band of carriers.  In this
coherent, heavy fermion state there are substantial antiferromagnetic
spin fluctuations which can be studied in great detail by magnetic
neutron scattering from single crystals.  This has been the topic of a
recent review (Aeppli and Broholm, 1994).  The present paper presents
highlights of some experiments which have occurred since then. 

\subsection{Neutron Scattering Cross Section}
Because of its magnetic moment the neutron can couple to moments in solids
via the dipolar force.  The energy and wavelengths of thermal and subthermal
neutrons are well matched to the energy and length scales of most condensed
matter systems and this is particularly true for heavy fermions.  We
will briefly review the formalism which describes the magnetic neutron
scattering.  For a
detailed treatment of the neutron scattering cross-section there are some
excellent texts which can serve as an introduction (Squires, 1978) or
more comprehensive exposition (Lovesey, 1984).  

The partial differential cross section for magnetic neutron scattering,
which measures the probability of scattering per solid angle per unit energy, 
is
\begin{equation}
\frac{d^2\sigma}{d\Omega dE'}=\frac{k'}{k}\frac{N}{\hbar}(\gamma r_o)^2
\left| f(\bf{Q}) \right|^2 \sum_{\alpha\beta} (\delta_{\alpha\beta}-
\hat{Q}_{\alpha}\hat{Q}_{\beta})S^{\alpha\beta}(\bf{Q},\omega)
\end{equation}
where $k (k')$ is the incident (scattered) neutron wavevector, N is the 
number of moments, $\gamma r_o = 5.391$ fm is the magnetic scattering length,
$f(\bf{Q})$ is the magnetic form factor (analogous to the electronic form 
factor appearing in the x-ray scattering cross section), $\bf{Q}$ is the 
momentum 
transfer, $\omega$ is the energy transfer, and the summation runs over the 
Cartesian directions.  $S^{\alpha\beta}(\bf{Q},\omega)$ is the 
magnetic scattering
function which is proportional to the space and time Fourier transform of
the spin-spin correlation function.  

If the incident and scattered neutron
energies are the same (elastic scattering) then the correlations at infinite
time are being probed and, in a magnetically ordered material, the
scattering function will contain delta functions at the wavevectors
corresponding to magnetic Bragg reflections.  The 
$(\delta_{\alpha\beta}-\hat{Q}_{\alpha}\hat{Q}_{\beta})$ term in the cross
section means that neutrons probe the components of spin perpendicular to
the momentum transfer, {\bf Q}.  If there is no analysis of the scattered
neutron energy then (within the static approximation) the measured intensity
is proportional to the Fourier transform of the instantaneous correlation 
function which is essentially
a snapshot of the spin correlations in reciprocal space.  At non-zero energy
transfers the spin dynamics of the system under study are probed.  In a
magnetically ordered system of localised spins the elementary magnetic 
excitations are spin waves.  

The fluctuation dissipation theorem relates the
correlations to absorption, in other words the scattering function is
proportional to the imaginary part of a generalised ({\bf Q} and $\omega$
dependent) susceptibility, $\chi''({\bf Q},\omega)$.  In the zero frequency, 
zero wavevector limit,
the real part of the generalised susceptibility is the usual DC susceptibility
measured by magnetisation. In a metal the elementary excitations are
electron-hole pairs.  Since it is possible to excite an electron-hole pair
by promoting a quasiparticle from below the Fermi surface to above the Fermi 
surface, and at the same time flipping its spin, neutrons can be used to
probe the low energy excitations of a metal.  The generalised susceptibility
(for a non-interacting metal) is just the Lindhard susceptibility which can
be calculated from the band structure.

\section{Antiferromagnetism and Superconductivity}
\subsection{$\rm{UPt_3}$}
\begin{figure}[tb]
  \begin{center}
    \epsfig{file=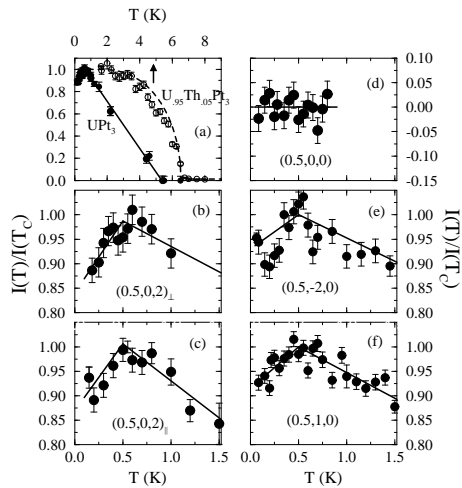,height=6cm}
  \end{center}
  \caption{Temperature dependence of the antiferromagnetic Bragg
peaks for ${\rm UPt_3}$.  (a)-(c) show the intensity measured with
x-rays (with neutron data for isostructural ${\rm U_{0.95}Th_{0.05}Pt_3}$
shown in (a) [open circles] for comparison).  (d)-(f) show the neutron
scattering intensity for three different Bragg reflection entering the
superconducting state. From Isaacs et al (1995).}
  \label{xrays}
\end{figure}
${\rm UPt_3}$ has remained a very popular  system because it
is both the quintessential strongly renormalized Fermi liquid, as revealed
especially by de Haas van Alphen experiments, and the quintessential
unconventional superconductor, displaying an array of properties ranging from
multiple superconducting phases to anisotropies not likely predicated on
normal state anisotropies. While the broad outlines of the ${\rm UPt_3}$ 
problem
were clear several years ago, the past two years have witnessed 
scattering experiments which have answered important outstanding questions.
These experiments all have to do with the weak antiferromagnetic order
whose Bragg signal is reduced by passing into the superconducting state,
and which is greatly enhanced- while superconductivity is 
eliminated- upon Th substition for U or Pd
substitution for Pt (Ramirez et al, 1986, de Visser et al, 1986,
Goldman et al, 1986 , and Frings et al, 1987). 
In particular, Isaacs et al (1995) performed a combination 
of X-ray and neutron diffraction experiments which showed the following
(see Figure \ref{xrays}):
(i) The reduction in the magnetic Bragg scattering found in an earlier
experiment (Aeppli et al, 1989)
is due to a reduction in the magnitude of the corresponding
magnetic moment, and not to rotation of the moments, e.g. in the basal
planes of the material.
(ii) There seems to be little difference between the behaviors of
the magnetic order exhibited in the near surface region probed by resonant
X-ray scattering and the bulk probed by neutrons.
(iii) The magnetic coherence length in quite heavily doped and 
non-superconducting ${\rm U_{0.95}Th_{0.5}Pt_3}$ is resolution-limited. 
This again 
emphasizes that a special local disorder is the most likely cause of
the magnetism in pure ${\rm UPt_3}$.

\begin{figure}[tb]
  \begin{center}
    \epsfig{file=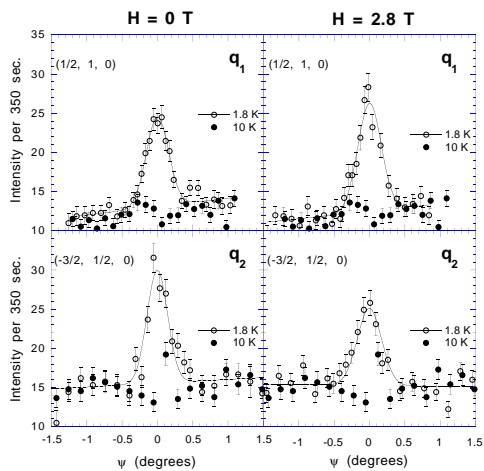,height=7cm}
  \end{center}
  \caption{Magnetic Bragg peaks for two different domains in $\rm{UPt_3}$ for
H=0 and 2.8 Tesla.  Complete selection of a single domain by the 2.8 T field
would eliminate the $\vec{q_2}$ Bragg peak and increase the $\vec{q_1}$ peak
by a factor of three. From Lussier et al (1996).}
  \label{domains}
\end{figure}
The second new scattering experiment also addressed the vector nature of
the ordered moment. In particular, Lussier et al (1996) investigated whether an
external magnetic field parallel to the basal planes- the 'easy' direction
as inferred from bulk measurements - could rotate the moments. A field of
up to 3.2 T was not able to either rotate the moments or select a single
domain (see Figure \ref{domains}). Given that
such a limiting field is beyond ${\rm H_{c2}}$ for the superconductor, finding
(i) of Isaacs et al (1995) is not surprising. Thus there are 
anisotropies, possibly
random, which strongly pin the small ordered moment in pure ${\rm UPt_3}$. It
will be interesting to see whether the same result obtains in the
more coherent antiferromagnetic state induced by Th and Pd impurities.
The finding that a single magnetic domain is not produced implies that
either the magnetic structure is not single-{\bf Q} or all measurements
of the superconducting phase diagram have been in multi domain samples,
requiring a re-evaluation of theories based on the symmetry breaking of
antiferromagnetic ordering.

\subsection{$\rm{UPd_2Al_3}$ and $\rm{UNi_2Al_3}$}
\begin{figure}    
  \begin{center} 
    \epsfig{file=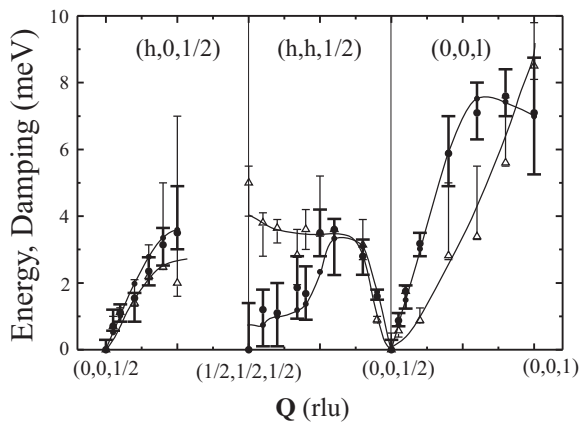,height=6cm,clip=%
                       bbllx=16,bblly=85,bburx=732,bbury=597} 
  \end{center}
  \caption{Wavevector dependence of the energy (filled circles) and damping
(open triangles) of the spin waves in the ordered state of $\rm{UPd_2Al_3}$.
There are well defined spin waves along the $\rm{c^*}$ axis however, in
the basal plane, the response is overdamped making it difficult to 
independently determine $\Gamma_{\bf Q}$ and $\omega_{\bf Q}$.}
  \label{dispersion}
\end{figure}
In 1991 two new heavy fermion compounds were discovered which displayed the
coexistence of antiferromagnetic order and superconductivity.  
$\rm{UPd_2Al_3}$
has an antiferromagnetic transition at 14.4 K and, in the best samples,
a superconducting T$_c$ of 2 K, the highest of any heavy fermion compound
at ambient pressure (Geibel et al, 1991a).  $\rm{UNi_2Al_3}$ has a somewhat 
lower T$_N$ (5.2 K) and T$_c$ (1 K) (Geibel et al, 1991b).  Both share the 
same hexagonal crystal structure (space group P6/mmm).

Powder neutron diffraction has shown that, in the antiferromagnetic state,
$\rm{UPd_2Al_3}$ has moments of 0.85 $\mu_B$ lying in the hexagonal basal
plane with the moments in a given layer ferromagnetically aligned and 
alternating up the c-axis (Krimmel et al, 1992a).  Initial reports of a
suppression of the ordered moment in the superconducting state by Krimmel
et al (1992a) have not been reproduced (Kita et al, 1994).  Measurements of
the magnetisation density in the paramagnetic state using polarized neutrons 
have shown that the magnetic moment resides on the U site with no spin
transfer to the Pd ions (Paolasini et al, 1993), comparison with 
magnetisation data suggest that there is an additional (12\% contribution)
from the polarisation of the conduction electrons.  A determination of the
magnetic phase diagram up to 5 T (Kita et al, 1994) has shown that the
moment lies along the a axis in the basal plane.  Application of a 
magnetic field perpendicular to one of the a axes 
(along $[\bar{1}10]$) favours that magnetic
domain and as the field is increased above a critical field of order
0.5 T the fraction of the sample with moments aligned along the a axis
perpendicular to H increases from 33\% to 100\%.  If the field is
applied parallel to [0,1,0] then then a two step process occurs: first
above 0.5 T the two domains at $\pi/3$ are selected, then above 4 T the
moments are constrained to lie perpendicular to the field along the
next nearest direction in the basal plane.  As the temperature is increased
towards ${\rm T_N}$ the fields for domain selection and reorientation
approach zero.

\begin{figure}
  \begin{center}
    \epsfig{file=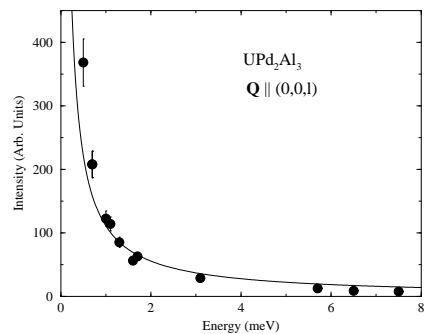,height=5cm}
  \end{center}
  \caption{Spin wave intensity as a function of energy for $\rm{UPd_2Al_3}$
obtained for momentum transfers displaced from $(0,0,\frac{1}{2})$ along the
c-axis.  The intensity is the amplitude for an inelastic Lorentzian response
convolved with the spectrometer resolution.  The line is the $1/\omega$ 
dependence expected for conventional spin waves.}
  \label{intensity}
\end{figure}
The inelastic neutron scattering from $\rm{UPd_2Al_3}$ has been studied
using powder, time-of-flight (Krimmel et al, 1996) and single crystal, 
triple-axis techniques (Petersen et al, 1994, Mason et al, 1995).
The powder measurements in the paramagnetic state show a strong
quasi-elastic response which is peaked at the wavevector corresponding to
the $(0,0,\frac{1}{2})$ Bragg peak.  The single crystal studies have shown
that in the antiferromagnetically ordered state this response evolves into 
spin waves which, within the limits of the experimental resolution of
0.35 meV, have no gap at the ordering wavevector.  The full dispersion
surface extracted from these measurements is shown in Figure \ref{dispersion}
along with the wavevector dependence of the spin wave lifetime.  These
quantities correspond to the energy and damping of an inelastic Lorentzian
response corrected for spectrometer resolution.  The structure of the
dispersion requires a model of the magnetic interactions with at least four 
groups of next neighbours implying long range interactions.  Moreover, it is
not possible to describe both energies and lifetimes in a localised moment
spin wave model (Lindg\aa rd et al, 1967) because damping arising 
from off-diagonal 
terms in the Hamiltonian results in
a zone centre gap inconsistent with the data.  This suggest the damping
is of extrinsic (conduction electron) origin.  The damping is generally 
comparable to the spin wave energy although for
wavevectors displaced along the c-axis there are well resolved modes with
a linear dispersion.  The intensity of the spin waves along the c-axis,
obtained from the same fits, is shown in Figure \ref{intensity} in comparison
with the $1/\omega$ expected for conventional spin waves.  Measurements of
the spin wave intensities in a single domain sample (produced as described
in the preceding paragraph) have shown the excitations are transverse
to the moment direction.  It appears that $\rm{UPd_2Al_3}$ is unique among 
U compounds in that it possesses conventional spin wave excitations with a 
very small or no gap at the ordering wavevector.  These spin waves are 
strongly 
damped due to interaction with the conduction electrons but, at energies
less than a few multiples of ${\rm k_B T_c}$, show no change on 
entering the superconducting
state (Petersen et al, 1994).  This is consistent with the results of
heat capacity (Caspary et al, 1993) and muon spin rotation measurements 
(Feyerherm et al, 1994) which have been
interpreted as evidence for two coexisting electronic systems, localised
5f magnetic states and delocalised states which are responsible for
superconductivity (Steglich et al, 1996).

\begin{figure}
  \begin{center}
     \framebox[5cm]{\rule[-.5cm]{0cm}{5cm}insert 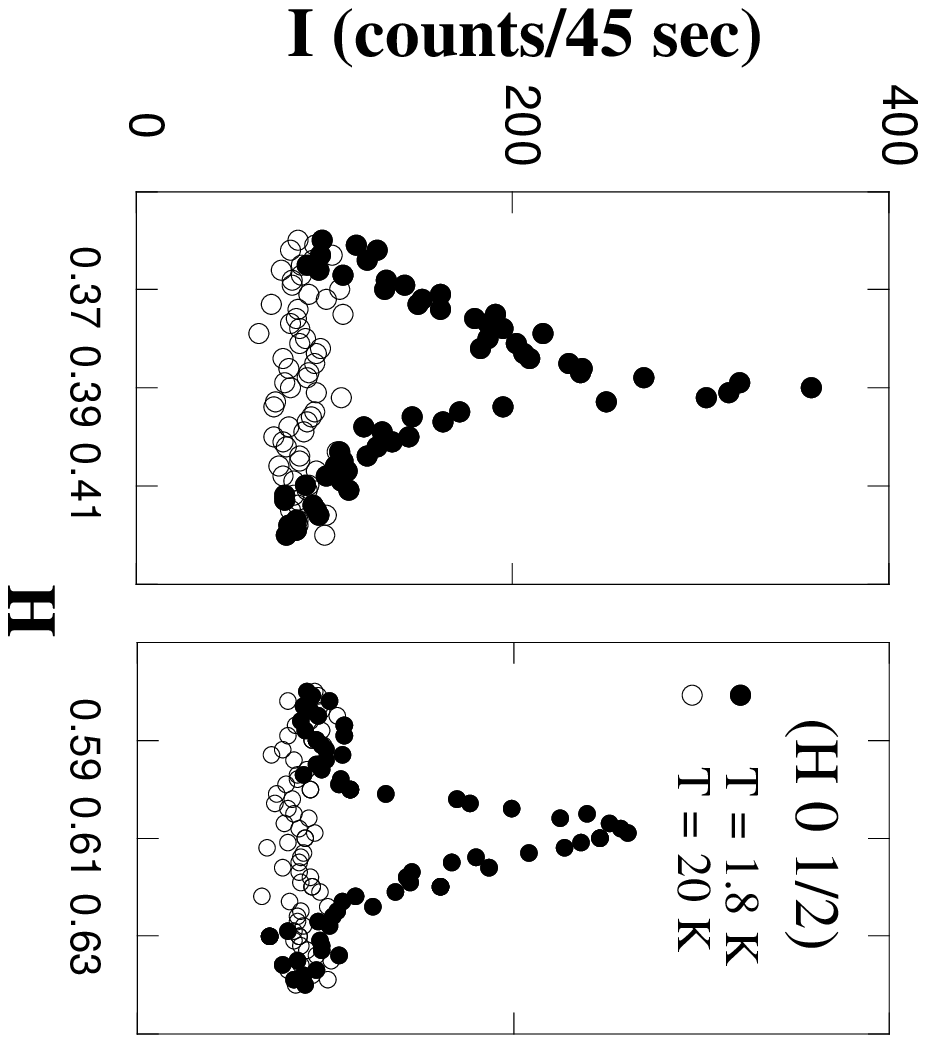}
  \end{center}
  \caption{Scans through the incommensurate peaks in $\rm{UNi_2Al_3}$
along the $(h,0,\frac{1}{2})$ direction above (open circles) and
below (closed circles) $\rm{T_N}\sim5.2 K$.  From Schr\"oder et al (1994)}
  \label{incommensurate}
\end{figure}
Initial powder diffraction studies of $\rm{UNi_2Al_3}$ failed to observe
any magnetic Bragg peaks below ${\rm T_N}$ and placed an upper bound on the
ordered moment of $0.2 \mu_{\rm B}$ (Krimmel et al, 1992b).  $\mu$SR 
experiments indicated that the ordered moment was likely of order 
$0.1 \mu_B$ (Amato et al, 1992).  Schr\"oder et al (1994) performed
neutron diffraction measurements on a single crystal of $\rm{UNi_2Al_3}$
and found that it ordered incommensurately below 5.2 K with an ordered
moment of $0.24 \pm 0.1 \mu_B$.  Figure \ref{incommensurate} shows scans 
through two of the incommensurate wavevectors,
$(\frac{1}{2}\pm\delta,0,\frac{1}{2})$ with $\delta=0.110\pm0.003$,
above and below ${\rm T_N}$.  The intensities of six magnetic Bragg peaks 
measured at 1.8 K are best described by a model structure which is a 
longitudinal spin density wave within the hexagonal basal plane with the
moments parallel to ${\rm a^*}$.  The moment direction in $\rm{UNi_2Al_3}$
is therefore
rotated $\pi/6$ compared to $\rm{UPd_2Al_3}$ but the observation of an
incommensurate modulation within the basal plane is perhaps not surprising
given the long range interactions manifested in the spin wave measurements
in $\rm{UPd_2Al_3}$.

\subsection{$\rm{UNi_4B}$}
\begin{figure}
  \begin{center}
    \epsfig{file=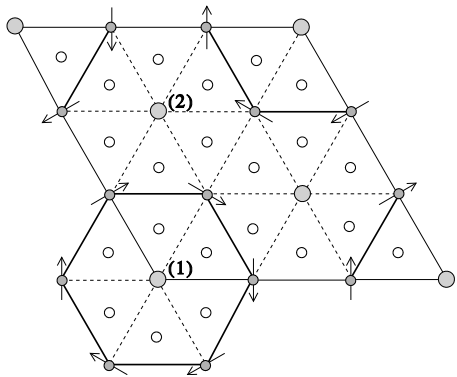,height=5cm,clip=,%
                bbllx=0,bblly=25,bburx=272,bbury=200}
  \end{center}
  \caption{Magnetic structure of $\rm{UNi_4B}$ projected onto the hexagonal
basal plane.  The magnetic layers are stacked ferromagnetically along the
c-axis.  The solid circles, labelled (1) and (2), represent the paramagnetic
U moments.  From Mentink et al (1994).}
  \label{pinwheel}
\end{figure}
One of the most intriguing new compounds which have been studied in recent
years is $\rm{UNi_4B}$ which has a structure based on the hexagonal 
${\rm CaCu_5}$ structure.  There is a small distortion which
modifies the local environment of $\frac{2}{3}$ of the U ions through
their collective motion towards the remaining sites (Mentink et al, 1996a).
As a result $\frac{1}{3}$ of the uranium moments are on six-fold
symmetric sites while the remainder are on two-fold symmetric sites. 
The resistivity, susceptibility, and specific heat of $\rm{UNi_4B}$ all
show anomalies typical for antiferromagnetic ordering at 21 K and this
has been confirmed by single crystal neutron diffraction 
(Mentink et al, 1994).  The magnetic structure, shown in Figure 
\ref{pinwheel}, is very unusual.  The moments on the outer, two-fold,
sites of hexagonal plaquets form a pinwheel-like structure while the 
moments on the central six-fold sites, which are frustrated due to
the cancellation of interactions with nearest neighbours, do not order.
The moments are ferromagnetically aligned along the c-axis.

Immediately below the phase transition there is a significant increase
in the DC and AC susceptibility (Mentink et al, 1996a) which is quenched
by the application of a modest magnetic field ($<1$ T).  This is likely the 
signature of the ferromagnetically correlated chains on the non ordering
sites.  At low temperatures ($<2$ K), however, this effect is eliminated,
the resistivity passes through a maximum and  c/T increases dramatically
to over ${\rm 500 mJ/(mole K^2)}$ at 0.3 K (Mentink et al, 1996b).  
It appears that the frustration is
being alleviated by the the formation of heavy itinerant states without
a moment in the presence of the localised moments which order at 21 K.
This is similar to what occurs in ${\rm DyMn_2}$ (Nu\~{n}ez Regueiro and
Lacroix, 1994) and CeSb (Ballou et al., 1991) and is the consequence of
the combination of lattice frustration, a proximity to a magnetic-nonmagnetic 
transition and strong anisotropy.

\section{Semiconductors and Non-Fermi Liquids}
\subsection{$\rm{CeNiSn}$}
\begin{figure}[tb]
  \begin{center}
    \epsfig{file=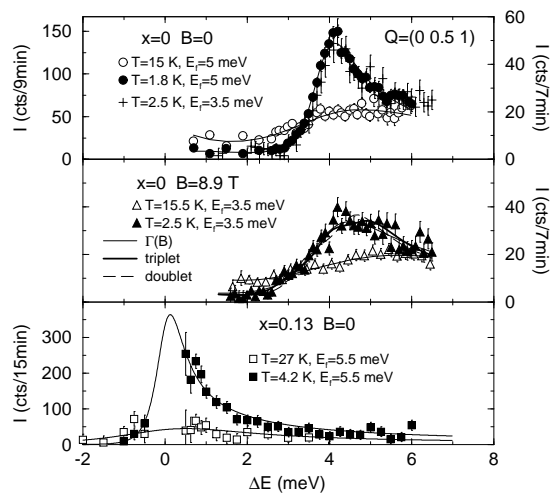,height=7cm}
  \end{center}
  \caption{Constant-{\bf Q} scans in $\rm{CeNiSn}$ for (upper panel) 
H=0, (middle panel)
H=8.9 T and (lower panel) 
$\rm{CeNi_{0.87}Cu_{0.13}Sn}$  at ${\rm {\bf Q}=(0,0.5,1)}$ showing the
effect of increasing temperature and magnetic field on the inelastic 
response.  From Schr\"oder et al (1996).}
  \label{conq}
\end{figure}
CeNiSn is interesting because it is the only Ce-based 'Kondo insulator'
(for a review see Aeppli and Fisk, 1992)
which can be readily fabricated in (large) single crystal form. 
Since the review of Aeppli and Broholm (1994), 
the material has received considerable attention from
various groups throughout the world. The principal new results are:\\
(i) The discovery of a clean gapped signal at wavevectors of type
(0,0.5,l) where l is an integer  in addition to those equivalent to (0,0,1)
(Kadowaki et al, 1994 and Sato et al, 1995).
Figure \ref{conq} shows the new peak, especially striking in its
much more intense manifestation after the new Ris\o~ cold neutron
guide tubes  were installed (for a comparison between this spectrum and 
that taken before the installation of the new guides see Lebech (1993)). 
While the gap is larger(4mev) at the former point than the latter (where it
is 2meV (Mason et al, 1992)), the property that 
$\chi''({\bf Q},\omega)$ is a strong 
function of {\bf Q}, while
$\chi'({\bf Q},\omega=0)$ is not, remains. T
Thus, the puzzle of the 'shielded' RKKY
interactions in Kondo insulators remains, although Varma (1995) has 
advanced arguments as to its resolution.\\
(ii) The discovery of long range magnetic order in CeNiSn samples
doped by Cu substitution for Ni  to achieve metallic heavy fermion behavior.
The magnetic ordering vector is close to the (0.5,l,0) vectors found
to exhibit the higher gap frequency. Thus, in addition to 
producing a (dirty) metal, 
doping apparently eliminates the shielding phenomenon
seen in the parent compound as well as the other celebrated single crystal 
Kondo insulator, FeSi.\\
(iii) The discovery that a magnetic field strongly affects the shape
of the magnetic gap spectra (see Figure \ref{conq}). 
In particular, the gap appears less 
sharp, although one cannot judge whether this is due to field-induced
splitting of some degeneracy or a true reduction in the lifetime of
excitations at the gap energy. In spite of the considerable spectral
change as well as the fact that the sample is rapidly approaching 
a metallic condition with increasing field, the shielding phenomenon
mentioned in (i) and (ii) remains. 
In summary, the most important consequence of the new work on CeNiSn
and its relatives is that there are dramatically different routes
to metallic behavior in heavy fermion systems, the first (doping) of which
leads
to substantial RKKY interactions  while the second (external field) does
not. 

\subsection{$\rm{Y_{1-x}U_xPd_3}$ and $\rm{UCu_{5-x}Pd_x}$}
\begin{figure}[tb]
  \begin{center}
     \framebox[7cm]{\rule[-.5cm]{0cm}{6cm}insert 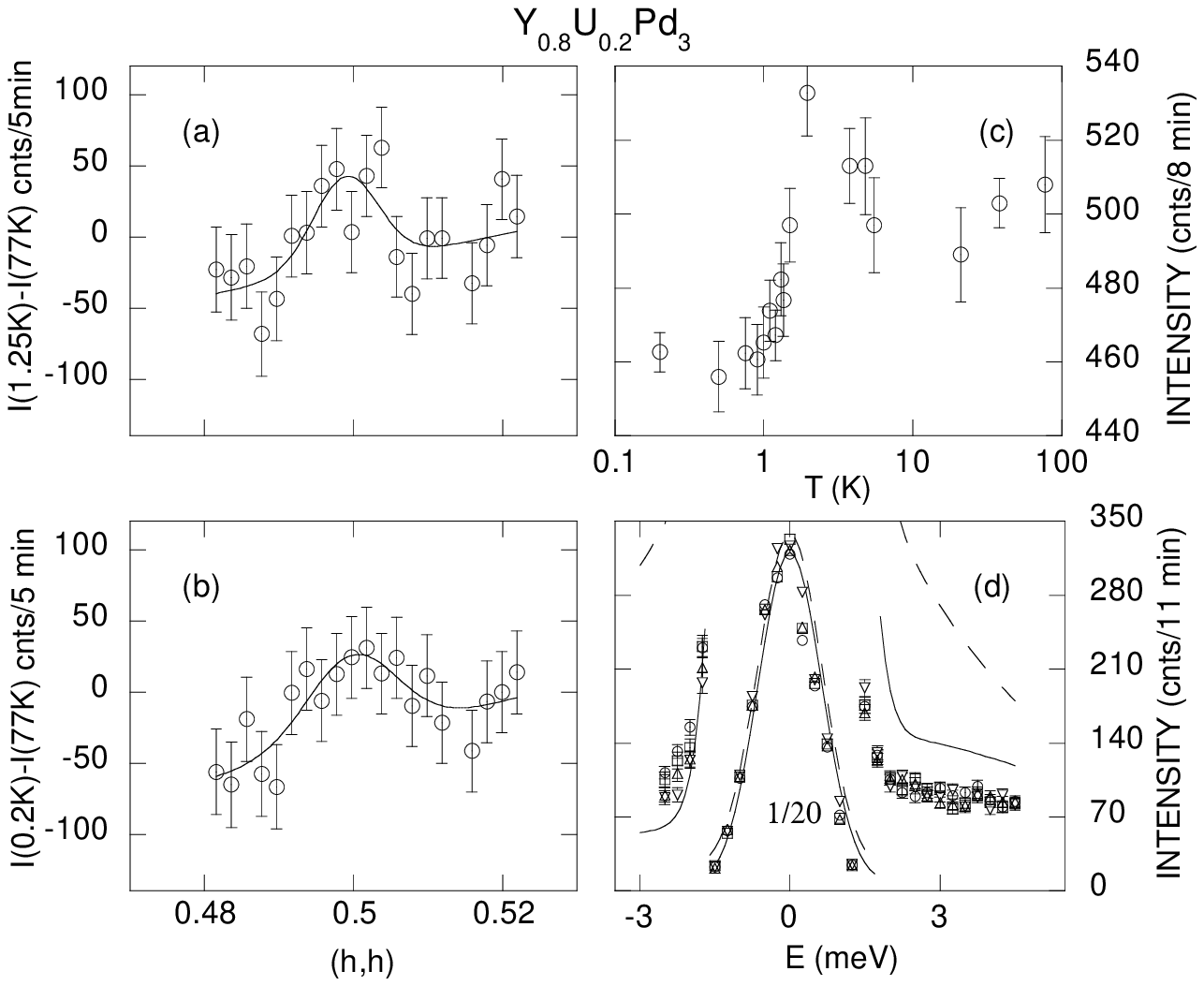}
  \end{center}
  \caption{Magnetic correlations in $\rm{Y_{0.8}U_{0.2}Pd_3}$. 
(a) {\bf Q} dependence of the energy integrated 
$S(\bf{Q},\omega)$ obtained by taking the difference in intensities at 1.25 K
and 77 K. (b) The same difference between 0.2 K and 77 K. (c) Temperature 
dependence of the scattering at 0.5 meV for {\bf Q} = (0.49,0.49,0). (d)
Constant-{\bf Q} scan at (0.5,0.5,0) at 70 K.  From Dai et al (1995)}
  \label{freezing}
\end{figure}
The properties of heavy fermion metals are a dramatic example of the
success of Fermi liquid theory in the sense that the low temperature
transport and thermodynamics, as well as the excitation spectra and
quantum oscillations in a magnetic field are all in accord with
predictions for a metal with a well defined Fermi surface (albeit with an
extremely large effective mass due to electronic interactions).
Similarly in Kondo insulators such as CeNiSn several distinct energy scales are
directly manifested in the size of the gap and the properties of
these materials are understandable in the framework of a band type
picture even though more careful examination of the neutron data reveal
important failings of band theory (Mason et al, 1992).  
There has been a great deal of interest recently in compounds,
typically random alloys, which exhibit weak power law and logarithmic
divergences in their low temperature properties at odds with the 
predictions of Fermi liquid theory, generically referred to as
non-Fermi liquid (NFL) behaviour.

One such material is $\rm{Y_{1-x}U_xPd_3}$ with x=0.2 (Seaman et al, 1991,
Andraka and Tsvelik, 1991) which has a logarithmically diverging 
electronic specific heat below 20 K, a power law divergence of the
susceptibility, and a resisivity which varies as $(1-(T/T_o))^{1.13}$.
This behaviour has been attributed to a two channel
quadrupolar Kondo effect (Seaman et al, 1991), proximity to a novel
zero temperature phase transition (Andraka and Tsvelik, 1991) or the
suppression of the Kondo temperature due to disorder and the 
associated proximity to a metal insulator transition (Dobrosavljevi\'{c}
et al, 1992).  Recent neutron scattering measurements by Dai et al (1995)
on polycrystalline samples with x=0.2 and 0.45 have shed considerable 
light onto the ground state for this material.
Figure \ref{freezing} summarizes some of the results.  Panels (a) and (b)
show the weak peak in the energy integrated cross section which develops at
low temperatures at the same antiferromagnetic wavevector at which long 
range order develops in the x=0.45 compound (which had previously thought 
to be a spin glass).  A temperature scan at 0.5 meV (panel (c)) shows
a suppression of these fluctuations as the characteristic energy moves 
to lower energies below about 2 K.  If the logarithmic increase in the
resistivity in this material were due to the conventional Kondo effect then
a quasielastic peak with a characteristic energy of 
${\rm k_B T \sim 3.6 meV}$ would result in a constant-{\bf Q} response
shown as solid and dashed lines in panel (d), inconsistent with the data.
Polarized beam measurements have shown that the dominant contribution to
the magnetic scattering for both the x=0.2 and x=0.45 samples is a
resolution limited response centred on zero frequency.  This indicates
that in both cases the ground state for the U ions is the $\Gamma_5$
triplet.  This magnetic ground state, suggested by the observation of
weak critical scattering, rules out the quadrupolar two-channel
mechanism for NFL behaviour in $\rm{Y_{1-x}U_xPd_3}$.
\begin{figure}
  \begin{center}
    \epsfig{file=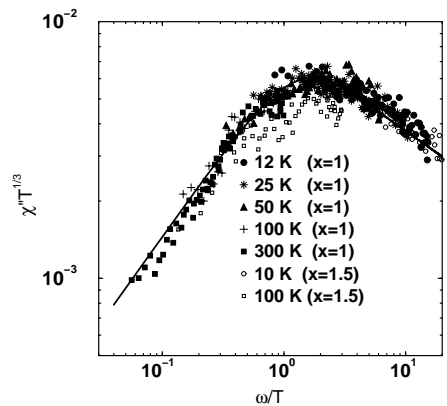,height=6cm}
  \end{center}
  \caption{$\rm{UCu_{5-x}Pd_x}$ exhibits scaling for both $x=1$ and 1.5.
The solid line corresponds to 
$\chi''(\omega,T)T^{1/3} \sim (T/\omega)^{1/3}tanh(\omega/1.2T)$.  From 
Aronson et al (1995).}
  \label{scaling}
\end{figure}

Another instance of NFL behaviour occurs in $\rm{UCu_{5-x}Pd_x}$.  In this 
case there is randomness due to alloying as in $\rm{Y_{1-x}U_xPd_3}$
however there is no site dilution of the U.  The novel low temperature
behaviour observed for x=1.5 has been ascribed to the suppression of a
spin glass transition to T=0.  Using time-of-flight powder measurements
Aronson et al (1995) have observed a magnetic excitation spectrum which,
below a cross over of about 25 meV, is characterised by a scale which is
determined by the temperature.  At energies lower than T the dynamic
susceptibility is proportional to $\omega/{\rm T}$, exactly cancelling
the temperature factor and leading to a temperature independent 
cross section, $S(\omega)$ similar to what has been seen in lightly
doped cuprates (Hayden et al, 1991).  This scaling behaviour, which is the
same for x=1 and 1.5, is explicitly shown in Figure \ref{scaling}.
Surprisingly for a dense Kondo lattice, there is no observable
Q dependence other than the overall form factor dependence.  This
could suggest a single ion origin for the observed scaling although it may
also be due to the directional averaging which occurs in any powder
measurement.  As in $\rm{Y_{1-x}U_xPd_3}$, the quasielastic 
response indicates a magnetic ground state which has an instability
driven towards T=0. 

The novel effects seen in these materials are not limited to alloys
with compositional disorder.  Similar effects are seen in 
${\rm URh_2Ge_2}$ (S\"{u}llow et al, 1996) although substitutional
disorder between Rh and Ge likely plays a role.  In that case there is
clearly a competition between spin glass and antiferromagnetic order
which may drive the low temperature properties.

\section*{Acknowledgements}
\addcontentsline{toc}{section}{Acknowledgements}
We would like to acknowledge the invaluable assistance of our many colleagues
who have participated in some of the experiments described in this review.
We also thank the authors of the papers listed in the Figure
captions for providing Figures for incorporation in this review and
W.J.L. Buyers for helpful suggestions.

\section*{References}

\addcontentsline{toc}{section}{References}

\begin{list}{}
{\leftmargin 20pt
\labelwidth 0pt
\itemsep 0pt\parsep0pt
\itemindent -20pt}

{\footnotesize

\item Aeppli G, and Broholm C, 1994:
{\it Handbook on the Physics and Chemistry of the Rare Earths}, {\bf 19}, eds.
Gschneidner, KA, Eyring, L, Lander, GH, and 
Choppin GR (Elsevier, Amsterdam) 123

\item Aeppli G, and Fisk Z, 1992: Comments Cond. Mat. Phys. {\bf 16} 155

\item Aeppli G, Bishop D, Broholm C, Bucher E, Siemensmeyer K, Steiner M,
and St\"{u}sser N, 1989: Phys. Rev. Lett. {\bf 63} 676

\item Amato A, Geibel C, Gygax FN, Heffner RH, Knetsch E, MacLaughlin DE,
Schank C, Schenk A, Steglich F, and Weber M, 1992: Z. Phys. B {\bf 86} 159

\item Andraka B and Tsvelik A, 1991: Phys. Rev. Lett. {\bf 67} 2886

\item Aronson MC, Osborn R, Robinson RA, Lynn JW, Chau R, Seaman
CL, and Maple MB, 1995: Phys. Rev. Lett. {\bf 75} 725

\item Ballou R, Lacroix C, and Nu\~{n}ez Regueiro MD, 1991: Phys. Rev.
Lett. {\bf 66} 1910

\item Caspary R, Hellmann P, Keller M, Sparn G, Wassilew C, K\"ohler R,
Geibel C, Schank C, Steglich F, and Phillips NE, 1993: Phys. Rev. Lett.
{\bf 71} 2146

\item Dai P, Mook HA, Seaman CL, Maple MB, and Koster JP, 1995:
Phys. Rev. Lett. {\bf 75} 1202

\item Dobrosavljevi\'{c} V, Kirkpatrick TR, and Kotliar G, 1992:
Phys. Rev. Lett. {\bf 69} 1113

\item Feyerherm R, Amato A, Gygax FN, Schenk A, Geibel C, Steglich F, Sato N,
and Komatsubara T, 1994: Phys. Rev. Lett. {\bf 73} 1849

\item Frings PH, Renker B, and Vettier C, 1987:
J. Magn. Magn. Mat. {\bf 63-64} 202 

\item Geibel C, Schank C, Theiss S, Kitazawa H, Bredl CD, 
B\"ohm A, Rau M, Grauel A, Caspary R, Helfrich R, Ahlheim U,
Weber G, and Steglich F, 1991a: Z. Phys. B {\bf 84} 1

\item Geibel C, Thies S, Kaczororowski D, Mehner A, Grauel A,
Seidel B, Ahlheim U, Helfrich R, Petersen K, Bredl CD, and
Steglich F, 1991b: Z. Phys. B {\bf 83} 305

\item Goldman AI, Shirane G, Aeppli G, Bucher E, and Hufnagl J, 1987:
Phys. Rev. B {\bf 36} 8523

\item Grewe N and Steglich F, 1991:
{\it Handbook on the Physics and Chemistry of the Rare Earths}, {\bf 14}, eds.
Gschneidner KA and Eyring L (Elsevier, Amsterdam) 343

\item Hayden SM, Aeppli G, Mook H, Rytz D, Hundley MF, and Fisk Z, 1991:
Phys. Rev. Lett. {\bf 66} 821

\item Holland-Moritz E and Lander GH, 1994: 
{\it Handbook on the Physics and Chemistry of the Rare Earths}, {\bf 19}, eds.
Gschneidner KA, Eyring L, Lander GH, and 
Choppin GR (Elsevier, Amsterdam) 1

\item Isaacs ED, Zschack P, Broholm CL, Burns C, Aeppli G, Ramirez AP,
Palstra TTM, Erwin RW, St\"{u}cheli N, and Bucher E, 1995: Phys.
Rev. Lett. {\bf 75} 1178

\item Kadowaki H, Sato T, Yoshizawa H, Ekino T, Takabatake T, Fuji H,
Regnault LP, and Isikawa Y, 1994: J. Phys. Soc. Jap. {\bf 63} 2074

\item Kita H, D\"onni A, Endoh Y, Kakurai K, Sato N, and Komatsubara T,
1994: J. Phys. Soc. Japan {\bf 63} 726

\item Krimmel A, Fischer P, Roessli B, Maletta H, Geibel C, Schank C,  
Grauel A, Loidl A and Steglich F, 1992a: Z. Phys. B {\bf 86} 161

\item Krimmel A, Loidl, A., Eccleston R, Geibel C, and Steglich F, 1996:
J. Phys.:Condens. Matter {\bf 8} 1677

\item Lebech B, 1993: Neutron News {\bf 4} 31

\item Ling\aa rd PA, Kowalska A, and Laut P, 1967: J. Phys. Chem.
Solids {\bf 28} 1357

\item Lovesey SW, 1984:
{\it Theory of Neutron Scattering from Condensed Matter}, (Clarendon Press,
Oxford)

\item Lussier B, Taillefer L, Buyers WJL, Mason TE, and Petersen
T, 1996: Phys. Rev. B submitted

\item Mason TE, Aeppli G, Ramirez AP, Clausen KN, Broholm C,
St\"{u}cheli N, Bucher E, and Palstra TTM, 1992: Phys. Rev. Lett.
{\bf 69} 490

\item Mason TE, Petersen T, Aeppli G, Buyers WJL, Bucher E, Garrett JD,
Clausen KN, and Menovsky AA, 1995: Physica B {\bf 213\&214} 11

\item Mentink SAM, Drost A, Nieuwenhuys GJ, Frikkee E, Menovsky 
AA, and Mydosh JA, 1994: Phys. Rev. Lett. {\bf 73} 1031

\item Mentink SAM, Mason, TE, Drost A, Frikkee E, Becker B, Menovsky AA,
and Mydosh JA, 1996a: Physica B in press

\item Mentink SAM, Amitsuka H, de Visser A, Slani\u{c} Z, Belanger DP,
Neumeier JJ, Thompson JD, Menovsky AA, Mydosh JA, and Mason TE, 1996b:
Physica B in press

\item Nu\~{n}ez Regueiro MD and Lacroix C, 1994: Phys. Rev. B {\bf 50} 16063

\item Paolasini L, Paix\~{a}o JA, Lander GH, Delapalme A, Sato N,
and Komatsubara T, 1993: J. Phys.: Condes. Matter {bf 47} 8905

\item Petersen T, Mason TE, Aeppli G, Ramirez AP, Bucher E, and Kleiman RN,
1994: Physica B {\bf 199\&200} 151

\item Ramirez AP, Batlogg B, Cooper AS, and Bucher E, 1986:
Phys. Rev. Lett. {\bf 57} 1072

\item Sato TJ, Kadowaki H, Yoshizawa H, Ekino T, Takabatake T, Fuji H,
Regnault LP, and Isikawa Y, 1995: J. Phys.: Condens. Matter {\bf 7} 8009

\item Seaman CL, Maple MB, Lee BW, Ghamaty S, Torikachvili MS, Kang J-S,
Liu LZ, Allen JW, and Cox DL, 1991: Phys. Rev. Lett. {\bf 67} 2882

\item Schr\"oder A, Lussier JG, Gaulin BD, Garrett JD, Buyers WJL, Rebelsky
L, and Shapiro M, 1994: Phys. Rev. Lett. {\bf 72} 136

\item Schr\"oder A, Aeppli G, Mason TE, St\"ucheli N, and Bucher E,
1996: to be published

\item Steglich F, Geibel C, Modler R, Lang M, Hellman P, and Gegenwart P,
1996: {\it Proc. Int. Euorconf. on Magnetic Correlations, Metal-insulator
Transitions and Superconductivity}; J. Low Temp. Phys. (in press)

\item Squires GL, 1978:
{\it Introduction to the Theory of Thermal Neutron Scattering}, (Cambridge
University Press, Cambridge)

\item S\"{u}llow S, Mentink SAM, Mason TE, Buyers WJL, Nieuwenhuys GJ,
Menovsky AA, and Mydosh JA, 1996: Physica B submitted

\item de Visser A, Klaase SCP, van Sprang M, Franse JJM, van den
Berg J, and Nieuwenhuys GJ, 1986: Phys. Rev. B {\bf 34} 8168
}

\end{list} 
\end{document}